\documentclass[conference,twocolumn]{IEEEtran}
\pdfoutput=1
\usepackage{amsopn,amsmath,amssymb,amsfonts}
\usepackage{cite}
\usepackage{graphicx,subfigure}

\graphicspath{{.}{./Results/}}

\title{Repeat-Accumulate Codes for Reconciliation in Continuous Variable Quantum Key Distribution}

\author{\IEEEauthorblockN{Sarah J. Johnson, Vikram A. Chandrasetty }\IEEEauthorblockA{School of Electrical Engineering and Computer Science\\ University of Newcastle, NSW, Australia.}
\and
\IEEEauthorblockN{Andrew M. Lance}
\IEEEauthorblockA{QuintessenceLabs Pty. Ltd., Canberra ACT  Australia.}}

\begin{document}

\pagestyle{empty}
\maketitle

\begin{abstract}
This paper investigates the design of low-complexity error correction codes for the verification step in continuous variable quantum key distribution (CVQKD) systems. We design new coding schemes based on quasi-cyclic repeat-accumulate codes which demonstrate good performances for CVQKD reconciliation.
\end{abstract}

\maketitle

\section{Introduction}

Quantum key distribution (QKD) is one of the most advanced applications of quantum physics and information science that enables the distribution of information-theoretically secure random key material between two parties in spatially-separated locations \cite{Gisin_2002}. Optical link configurations include optical fibre as well as free space links between ground, airborne and satellite nodes.
Quantum key distribution has advanced from a subject of pure research since its discovery in the mid 80's to being commercialized by several companies and has been demonstrated in various deployments: metropolitan quantum networks, long-distance free space links, securing banking transactions, and the transmission of election ballot results.

There are two complementary approaches to quantum key distribution: discrete variable quantum key distribution uses single-photon or weak coherent states and single photon detectors \cite{Scarani_2009_RevModPhys_QKD}, while continuous variable quantum key distribution (CVQKD), uses coherent states and homodyne detectors \cite{Weedbrook_2012}. Both discrete and continuous quantum key distribution systems have been experimentally demonstrated (for a review see \cite{Lo_2014}) and importantly both the discrete and continuous approaches to quantum key distribution have been proven to be information-theoretic secure, the later against collective attacks \cite{Garcia-Patron_2006,Navascues_2006}.

Continuous variable quantum key distribution has a gained interest recently because of the potential technology advantage it offers that may enable higher secret key rates. Technological advantages include: homodyne detectors, which exhibit  higher quantum efficiencies compared with single photon detectors; high-speed commercial-off-the-shelf components including modulators and detectors with bandwidths exceeding 10GHz; compatibility with optical networks; and detectors that can operate in bright-light conditions for twenty-four hour free-space optical link applications.

CVQKD protocols involve encoding random information onto the quadratures of coherent states of light, which are transmitted over an optical channel and measured using homodyne detectors. Both the sender and transmitter subsequently perform a series of post-processing steps on their respective raw key material, potentially including: key-sifting, post-selection, error reconciliation and privacy amplification \cite{Weedbrook_2012}. The secret key rate of CVQKD systems can be limited by the real-time processing speeds of the post-processing steps. In particular, error reconciliation can be particularly computationally intensive.

Modern CVQKD system employ forward-error correction (FEC) codes to limit the information leakage to any potential eavesdropper and hence maximize secret key rates. The more efficient the code (how closely it performs to the Shannon limit) the less information leaked. Currently, the efficiency and throughput of error correction coding is the limiting factor for the range of CVQKD systems \cite{Leverrier_2009}.
Operation at low signal-to-noise ratios is essential for security, and, the low rate iterative error correction codes required for very low signal-to-noise ratio (SNR) channels are not well studied in the error correction literature. Indeed, typical communications systems are designed to ensure the SNR is reasonable, for example by increasing the transmission power, and so very low rate error correction codes have not been required for traditional communications applications.

Recently, researcher have begun to consider how to design very low rate error correction codes for CVQKD including layered low-density parity-check codes \cite{Lodewyck2007_LDPC_CVQKD,Fossier_CVQKD_NJP2009}, multi-edge-type low-density parity-check codes \cite{Jouguet-2011-multiedge} and polar codes \cite{Jouguet_Polar_2012}. It has been shown that layered low-density parity-check codes with a block length of 200000 bits were sufficient to achieve efficiencies of 85\% over a range of SNRs \cite{Lodewyck2007_LDPC_CVQKD,Fossier_CVQKD_NJP2009,Jouguet2012_experimental}. However, high efficiency has only been achieved with relatively large block length (approximately 200 Kbits) and randomly constructed codes  \cite{Lodewyck2007_LDPC_CVQKD}, which make hardware implementations unrealistic. While the optical transmission of CVQKD offers high data rates, the demanding computing power of the error correction step reduced key rates to just over 2kb/sec on a 25km optical link \cite{Lodewyck2007_LDPC_CVQKD} or 8kb/sec on a 15km optical link \cite{Fossier_CVQKD_NJP2009}. Practical reasons for the decrease of the secret key generation rate are driven primarily by the impossibility of post processing all the data at the optical emission rate at high SNRs and a limited reconciliation efficiency at low SNRs \cite{Fossier_CVQKD_NJP2009}. In another scheme, CVQKD systems have been demonstrated with very high efficiencies ( $\geq 95\%$) using low rate multi-edge-type low-density parity-check (LDPC) codes with block lengths of approximately 1Mbit  \cite{Jouguet-2011-multiedge}. However, a quite high word error rate (1 in 3 codewords discarded) limits the overall key rate and the implementation of these long irregular codes in hardware has not been attempted. This paper investigates the design of low-complexity high-speed repeat-accumulate error correction codes for the verification step in continuous variable quantum key distribution using commercially proven code structures, namely quasi-cyclic repeat-accumulate codes. Our motivation is low cost hardware implementation of codes suitable for the CVQKD application. Of all the existing, commercially available, codes, the length 64,800 bits, DVB-S2 quasi-cyclic repeat-accumulate (RA) codes \cite{ETSI-DVBS2-2009} have parameters most suited to the CVQKD application, and the motivation for investigating this type of code is the availability of extremely fast commercial ASIC implementations.

The remainder of the paper is structured as follows:
Section~\ref{sec:Background} introduces CVQKD and describes RA codes.
Section~\ref{sec:NewCodes} designs new coding schemes for the CVQKD application and details results of the secret key rate of a CVQKD system using the designed repeat accumulate codes.

\section{Background} \label{sec:Background}

\subsection{Continuous Variable Quantum Key Distribution}

In CVKQD the final secret key rate can be expressed by the formula (see e.g. \cite{Lodewyck2007_LDPC_CVQKD}):
\vspace{-0.5em}
\begin{equation} \label{eq:key_rate}
\Delta I = (\beta I_{\rm AB} - I_{\rm E})(1-p_\mathrm{fail}).
\vspace{-0.5em}
\end{equation}
where $I_{\rm AB}$ denotes the mutual information between sender and receiver, $I_{\rm E}$ denotes the bound on an eavesdropper's maximum accessible information, $\beta$ denotes the error correction code efficiency and $p_\mathrm{fail}$ is the rate at which the decoder fails to decode (i.e. the word error rate).

To achieve high key generation rates and longer transmission distances for a CVQKD system, it is desirable in the error reconciliation step to correct errors close to the theoretical limits to minimize the information exposed to a potential eavesdropper (i.e. $\beta \rightarrow 1$) while also maintaining a reasonable word error rate and high throughput.



It was recently shown that if the sender encodes with Gaussian signals, a mapping protocol can be used to transform the Gaussian symbols to binary symbols for subsequent error correction \cite{Leverrier_2008}. The losses from this mapping are small at the low signal-to-noise ratios required for CVQKD reconciliation. Under this scheme, the receiver's measurements can be modeled by an additive white Gaussian noise (AWGN) channel with a signal-to-noise ratio (SNR) determined by the parameters of the CVQKD model \cite{Jouguet-2011-multiedge}.
Reconciliation is then performed using an error correction code designed for the binary-input (BI)-AWGN channel. The efficiency of the reconciliation code is measured by \cite{Jouguet-2011-multiedge}:
\begin{equation} \label{eqn:efficiency}
\beta(s) = \frac{R}{C(s)}
\end{equation}
for a given signal-to-noise ratio, $s$, where $R$ is the rate of the error correction code and $C(s)$ is the rate of the capacity-achieving code at this signal-to-noise ratio.

\subsection{Repeat-Accumulate Codes}

A repeat-accumulate code can be thought of as a type of low-density parity-check code with additional structure in the parity-check matrix to enable low-complexity encoding. An RA code can alternatively be seen as a type of serially concatenated turbo code (see \cite{Divsalar_Turbolike,Jin_IRA,Johnson_RA,Johnson-2010-Book} for more details). Briefly, a systematic RA code can be described by a parity-check matrix, $H$, that has two parts,
$H = [H_1,A]$.
Here $H_1$ is an $M \times K$ binary matrix, with column weights
$q_1, \ldots, q_K$, and row weights $a_1, \ldots, a_M$.
$A$ is an $M \times M$ matrix called an accumulator which
has the form:
\begin{equation} \label{A_RA}
A = \left[
  \begin{array}{cccccccccc}
     1 & 0  & 0 &   & 0 & 0  \\
     1 & 1  & 0 &   & 0 & 0  \\
     0 & 1  & 1 &   & 0 & 0  \\
       &  \vdots  &   & \ddots &  & \vdots  \\
     0 & 0  & 0 &      & 1 & 0  \\
     0 & 0  & 0 &   & 1 & 1  \\
 \end{array}
\right].
\end{equation}
The code rate is $R = K/N = (N-M)/N$ where the second equality is due to the structure of $A$ which ensures that $H$ is full rank. The structure of RA codes makes them systematic and suitable for fast encoding using back-substitution directly based on $H$. No generator matrix need be found. This gives them a significant advantage over standard LDPC and multi-edge-type LDPC for which an efficient generator matrix is not guaranteed.

One current application of RA codes can be found in the DVB-S2 standard for digital video broadcasting \cite{ETSI-DVBS2-2009}.
In the DVB-S2 RA codes the columns in matrix $H1$ can have one of two weights $q_1>3$ or $q_2 = 3$ where the fraction of the $K$ columns which are weight $q_1$ is given by $x$ and the fraction of the $K$ columns which are weight 3 is $1-x$. The values of $q_1$ and $x$ are different for each of the DVB-S2 codes (see \cite{ETSI-DVBS2-2009} for these values for each code). In order to ensure low-complexity hardware, the columns of $H_1$ are divided into sets of 360 columns where all of the 360 columns in the set are formed by cyclically shifting the previous column by a certain number of places. Thus to completely specify any given DVB-S2 code only the indices in the non-zero entries of $K/360$, instead of $K$, columns are required. This drastically reduces the memory required to store the set of DVB-S2 codes and allows fast encoding via shift registers and modulo-2 addition.

The DVB-S2 standard defines a class of eleven repeat-accumulate codes that span a range of code rates:
\[[1/4, 1/3, 2/5, 1/2, 3/5, 2/3, 3/4, 4/5, 5/6, 8/9, 9/10]. \]

\section{Reconciliation of Gaussian Variables using Repeat-Accumulate Codes} \label{sec:NewCodes}

\subsection{Designing a low rate RA code}

Because the lowest rate DVB-S2 code is only rate-1/4, we will also design a rate-1/10 RA code, which requires very little hardware modification over the existing code set to implement.

To design the lower rate DVB-S2-like RA code we will restrict our new code to have exactly the same structure as the existing DVB-S2 codes, and so the rate $R$ is restricted by
\[ RN =  0 \;\; \mathrm{modulo} \;\; 360, \]
and the fraction $x$ of the higher weight columns is restricted by
\[ xK  = 0 \;\;\mathrm{modulo} \;\; 360. \]

To find a rate-1/10 RA code with the same structure as the DVB-S2 codes, we used density evolution for RA codes modified to constrain the RA degrees to fit the DVB-S2 structure. Then a modified progressive edge growth algorithm was designed to find the coefficients for each column set. The parameters of the new rate-$1/10$ code are given in Appendix A.

\subsection{Efficiency and Complexity}

The failure rate of an error correction code is called its word error rate (WER). It is assumed that these failures to decode correctly are detected by the decoder, which is a reasonable assumption for belief propagation decoding of repeat-accumulate codes. An undetected error, which is theoretically possible, though rare, occurs if the belief propagation decoder converges to an incorrect codeword. However, these errors can easily be picked up by the addition of a very high-rate parity-check code concatenated with the repeat-accumulate code.

The efficiency of an error correction code is measured by comparing the code rate, $R$, of the error correction code to $C(s)$, the code rate of a theoretical capacity-achieving code at the same SNR. Determining which SNR corresponds to the error correction code can be calculated in one of two ways. Theoretically, for infinite length codes, by using density evolution to find the threshold (the SNR at which the word error rate goes to zero) of the ensemble of all possible codes with a given parameter set \cite{Richardson_irregLDPC}. Or, for finite-length codes, by simulation to measure the word error rate curve for a particular code as the SNR is varied. In real codes, the word error rate does not instantaneously drop to zero at a single threshold SNR value, rather it decreases more gradually as the SNR is increased. Consequently the measured SNR for a given code will depend on the word error rate which can be tolerated by the application. For example, the code rate-1/3 DVB-S2 code efficiency drops from $93$\% to $90$\% when the allowed WER is dropped from 1/2 to 1/100. This reduction in efficiency with WER is even more pronounced in lower rate codes. Consequently, it is critical when comparing efficiency to be clear about the word error rate being considered.

%
%

The implementation speed of repeat-accumulate codes varies with the technology used, the code rate and the channel noise. However, as the DVB-S2 codes have been specifically designed to enable high speed and low complexity implementation, decoding speeds can be very high. Reported decoding speeds are around 80-200 Mb/s on a GPU \cite{Falcao2011_GPU} and around 300 Mb/sec on an ASIC \cite{Muller2009_DVB2_ASIC}. Encoding speeds are much faster. By comparison, the two existing forward error correction schemes proposed for reconciliation in CVQKD report speeds of around 120kb/sec on a GPU, using multilevel coding with optimized irregular LDPC codes \cite{Lodewyck2007_LDPC_CVQKD}, and up to 7 Mb/s on a GPU for multi-edge type LDPC codes \cite{Jouguet2012_experimental}. Due to the code complexity and length (200Kb and 1Mb respectively) ASIC implementations have not been attempted.

A complicating factor is that decoder performance changes significantly with the number of decoder iterations performed. For the DVB-S2 codes the error rate is significantly affected by the maximum allowed number of iterations when this value varies from 10 to 100. Increasing the maximum number of iterations from 100 to 500, however, shows only a small improvement in performance. In general, decoding is halted early if a valid codeword is found and so the average number of iterations can be significantly less than the maximum.



\subsection{Operating over a continuous range of SNR}

A typical approach to operating over a continuous range of SNRs is to use a `code hopping' strategy to span the channel SNRs by switching between each of a set of codes as the channel varies \cite{Jouguet2012_experimental}. If a channel has a higher SNR than required for a particular rate code, but not high enough for acceptable performance of the code at the next rate up, the lower rate code is used. This results in a better error correction performance (in terms of an error rate lower than the specified minimum) but a lower efficiency because a lower rate code than is strictly necessary is used.

A more efficient strategy is to slightly increase or decrease the rate of the code, via puncturing or repetition coding, as the channel SNR increases or decreases \cite{Jouguet-2011-multiedge}. Puncturing is a commonly used strategy in error correction coding of not transmitting a selected set of bits in the codeword. The decoder acts as if those bits have been erased, which for the decoder amounts to setting the a priori probabilities for those bits to be a half (i.e. equally likely to be a 1 or 0). Since the number of message bits has not changed, but the number of transmitted bits has, the effective rate of the code has now been increased. The idea in repetition coding \cite{Leverrier_phd2009} is to use a sequence of $k$ received bits to transmit just one codeword bit (in effect a rate-$1/k$ repetition code) and in so doing raise the SNR for that combined bit to $k$ times the original SNR \cite{Leverrier_phd2009}. This strategy facilitates forward error correction coding in channels with SNRs lower than would otherwise be possible, but reduces the efficiency as the overall code rate is now $1/k$ times the error correction code rate.

In this work we also use puncturing to increase the code rate of the DVB-S2 codes, however, for lower rate codes we propose a code lengthening approach to improve on repetition coding at low SNRs. Starting with the RA parity-check matrix $H$ in \eqref{A_RA}, additional rows and columns are appended as shown in Figure~\ref{fig:ExtendedH}. The matrix $I$ is the identity matrix, (ones on the diagonal, zeros elsewhere) and the matrix $0$, is an all zero matrix. The matrix $E$ can be formed from the first two columns of a truncated Vandermonde matrix \cite{Benmayor_IRAextnd2008} or the parity-check matrix of a row-weight-2 LDPC code. This extended parity-check matrix, $H_\mathrm{etnd}$, is still systematic and easily encoded from $H_\mathrm{etnd}$ via back-substitution. Furthermore, the extra parity-bits can be generated independently from the original codeword and extra parity bits transmitted subsequently if the channel SNR is worse than expected.

\begin{figure}[th]
    \centering

    $H_\mathrm{etnd} =$ \large \begin{tabular}{|c|c|c|} 
    	\hline $H_1$ &  $A$\; & \;$0$\; \\
    	\hline  $E$\; & $0 $ & $I $ \\
    	\hline
    \end{tabular}
    \caption{Rate reduction via extending.
    \label{fig:ExtendedH}}
\end{figure}

\subsection{Efficiency results}

\begin{figure}[th]
    \centering
    \includegraphics[width=\columnwidth]{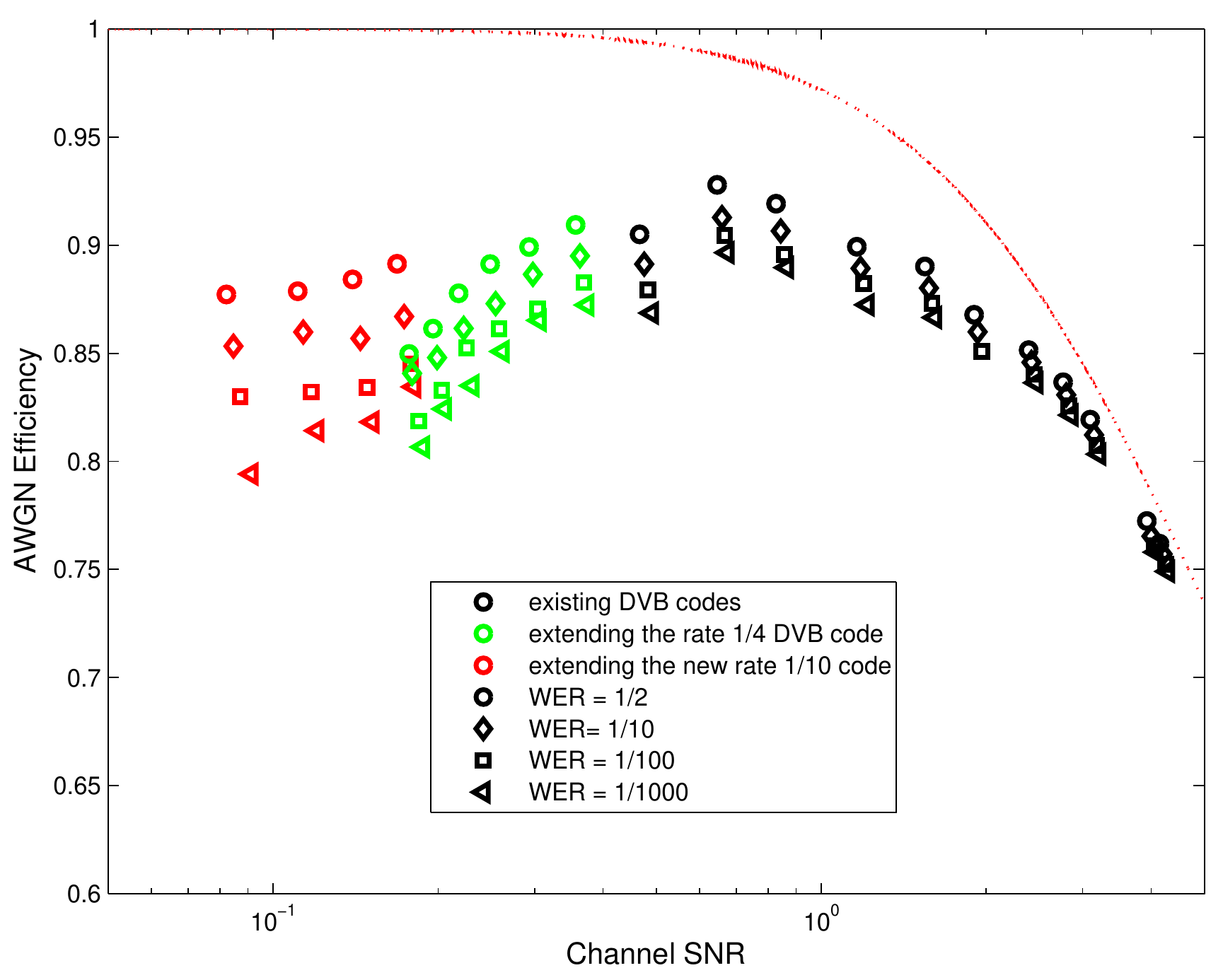}
    \caption{Efficiency of various codes for word error rates of $0.5$, $0.1$, $0.01$ and $0.001$. In black is the efficiency of the DVB-S2 codes, in green is the efficiency of codes formed by extending the rate-1/4 DVB-S2 code and in red is the efficiency of the new rate-1/10 code and new codes formed by extending it.
    \label{fig:DVB_HopExtnd_Efficiency_AWGN}}
\end{figure}

Figure~\ref{fig:DVB_HopExtnd_Efficiency_AWGN} shows the efficiency of the DVB-S2 codes for word error rates of $0.5$, $0.1$, $0.01$ and $0.001$ respectively. Also shown is the efficiency of codes formed by extending the lowest rate DVB-S2 code (rate-1/4) and the efficiency of the new rate-1/10 RA code and extensions.

\begin{figure}[th]
    \centering
    \includegraphics[width=\columnwidth]{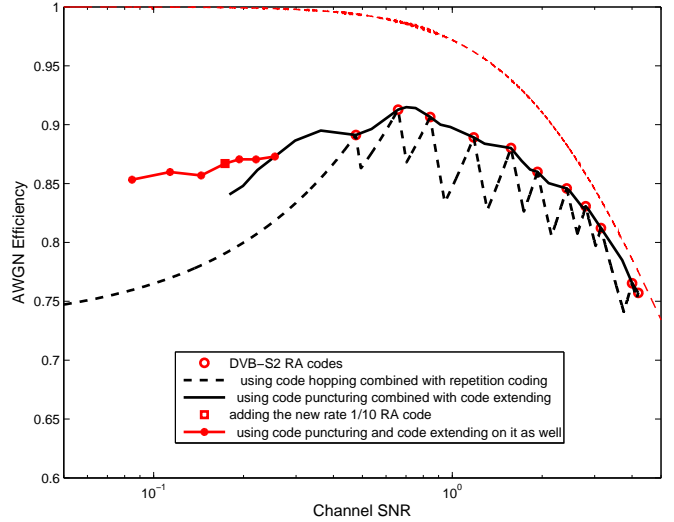}
    \caption{Efficiency, measured at WER = 1/10, of various coding schemes for the DVB-S2 codes.)
    \label{fig:DVB_HopExtnd_compr_AWGN}}
\end{figure}

Combining the DVB-S2 and DVB-S2-like RA codes with traditional repetition coding and code hopping gives the efficiency curve shown as a dashed line in Figure~\ref{fig:DVB_HopExtnd_compr_AWGN}. Applying puncturing and extending gives the efficiency curve shown as a solid line. Combining puncturing and extending with the new rate-$1/10$ DVB-S2-like code gives efficiencies above $0.85$\% over a wide range of SNRs. Higher efficiencies are possible if the allowed word error rate is increased above 1-in-10.

As can be seen in Figure~\ref{fig:DVB_HopExtnd_compr_AWGN} code extending can significantly improve the performance of repetition coding to reduce the code rate. Further, adding the new rate-1/10 RA code significantly improves efficiency in low SNR channels over using existing DVB-S2 codes alone. This is particularly critical in CVQKD applications where the efficiency of the codes in low SNR channels limits the distance of operation of the CVQKD scheme.

\subsection{Key rates in a CVQKD heterodyne system}

In this section we consider the performance of repeat-accumulate codes as the reconciliation step in a CVQKD system as described in \cite{Fossier_CVQKDopt_2009}.
We assume a coherent state CVQKD system with homodyne detection employing reverse reconciliation. The eavesdropper is assumed to utilize collective attacks and we assume asymptotic key length, i.e. finite key-size effects are not considered. (See \cite{Fossier_CVQKDopt_2009} for the derivation of the key rate equations.) 

\begin{figure}[th]
    \centering
    \includegraphics[width=\columnwidth]{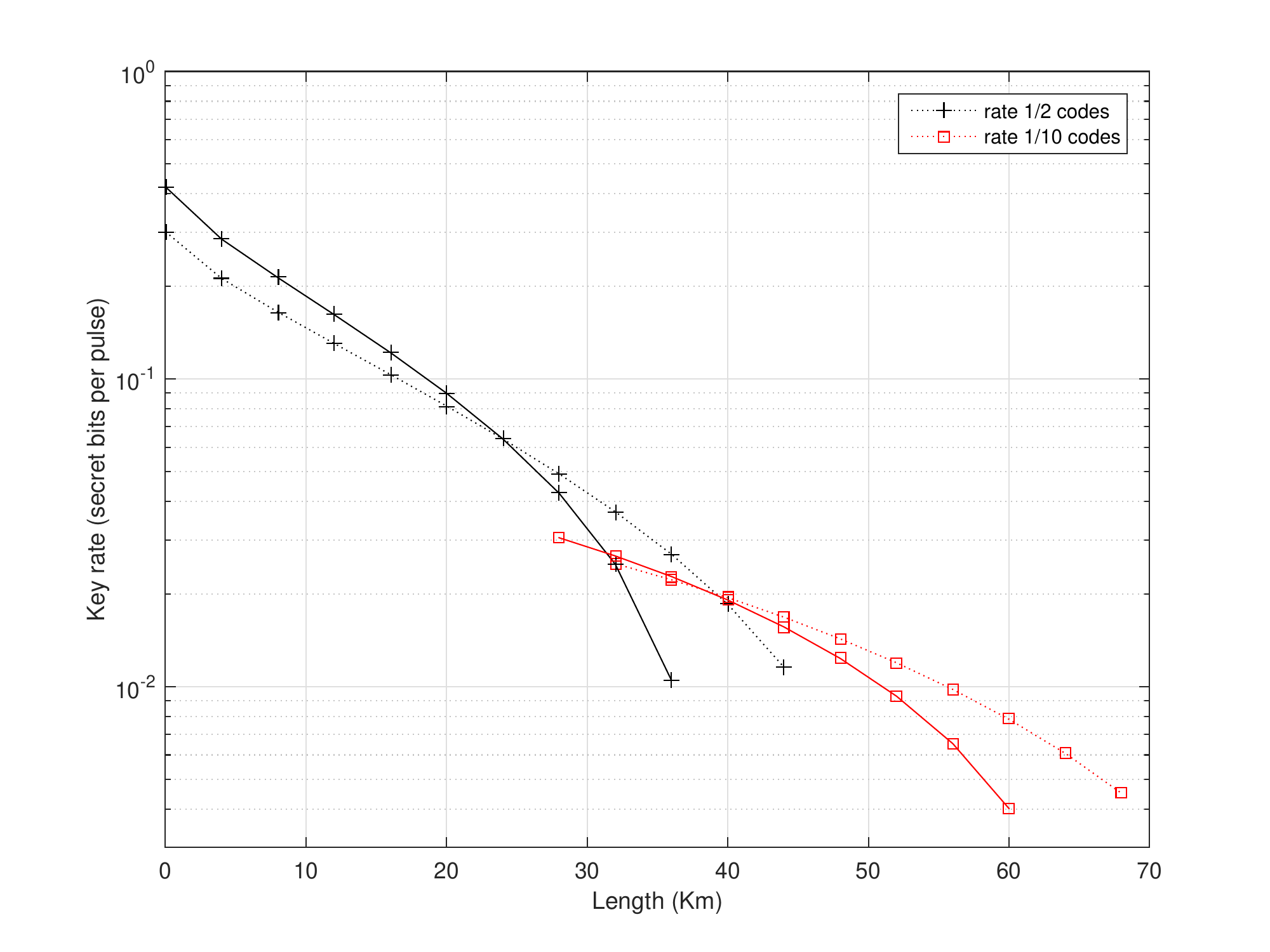}
    \caption{Secret key rate as a function of channel loss for a CVQKD system and with excess noise $\epsilon= 0.01$, detection efficiency $\eta = 0.6$, electronic noise $v_\mathrm{el} = 0.01$, and attenuation $0.2$ dB / Km. Solid curves show the quasi-cyclic RA codes (length $64,800$ rate $0.5$ and $0.1$) and dashed curves show the ME codes (length $2^{20}$ and rate $0.5$ and $0.1$) from \cite{Jouguet-2011-multiedge}.}
    \label{fig:DVB_KeyRate}
\end{figure}

The CVQKD model has the following free parameters: The encoded signal variance, $V_{\rm A}$, the channel properties of transmission, $T$, channel excess noise, $\epsilon$, homodyne detector efficiency, $\eta$, and electronic noise $v_{\rm el}$ at the receiver. In Figure~\ref{fig:DVB_KeyRate} we choose these parameters to replicate \cite[Figure 5]{Jouguet-2011-multiedge} and compare the rate $1/2$ and $1/10$ quasi-cyclic RA codes to the ME LDPC codes in \cite{Jouguet-2011-multiedge}. We see that the quasi-cyclic RA codes perform comparably to the multiedge LDPC codes despite being shorter, structured, and specifically designed to be efficient to implement.

\section{Conclusion}

Overall, quasi-cyclic repeat-accumulate codes modeled on the DVB-S2 codes can provide reasonably good key rates in a CVQKD heterodyne system compared to state of the art codes. Given their low implementation complexity, and existing high-speed ASIC implementations, this makes them a viable candidate for commercial CVQKD systems.


\section{Appendix A}

Description of the circulants in the $H_1$ part of the rate-1/10 RA code. Each row in the table gives the row indices of the first circulant column for each of the 18 column sets. Each column set contains 360 columns. The format follows that of the DVB-S2 codes. See \cite{ETSI-DVBS2-2009} for details on encoding methods.

\begin{small}
\begin{tabular}{l|l}
\# & Row Indicies \\
\hline
1 &     0  15960  45564  48399   2007  19137  45463  33191  39884  \\ & 45109  41483  19553  51001  33262  57015  26986  21998 \\ & 24833  42258 \\
2 &  2292  41854  46985  49651  52444   9757  53528    413  36760 \\ & 43848   42277  34280  32901  14138  54562  43304  46148 \\ &  9531    678 \\
3 & 21678   8451  45933   9635   7184  12580   8135  21014  23184 \\ & 19605   57842  17646  19672  34665  50644  26193  49109 \\ & 47297   2526 \\
4 &  1138  33012  34598  31489  56740  52609  15634  \\ & 7597   985  46  53850  19656  35877  44343  22954   25459 \\ &  57073  48417  15447 \\
5 & 26306   3957  35415  50885   8951  34488  14059  17762 \\ &  3933  799 36619   22981  27176  31670  24464   42482 \\ &  21818   8315    734 \\
6 & 37644  32896  30382  32399  13304  32643   8290  55214 \\ & 18273  36972  12505   48674  28470  47239  40769  36028  \\ &  5018  50298  17934 \\
7 & 25904  11332  18160  29161  12384  37865  25403  28531 \\ & 33394  2199  14952   13698  53705  40498  31923 26207 \\ &  23788  41053  48302 \\
8 & 34399  54407  52709 \\
9 & 21971  50797   7809 \\
10 & 51405  12545  14619 \\
11 & 24733  56388   6919 \\
12 & 25934  55070  14776 \\
13 & 35296  53121  34611 \\
14 & 33683  55690  50123 \\
15 &  1236  30161  10606 \\
16 & 14513  36456  24067 \\
17 &  4297  46488  32303 \\
18 & 20455   8354  55976 \\
\end{tabular}
\end{small}

\end{document}